\documentstyle[aps,epsbox,twocolumn]{revtex}
\begin{document}
%
%
\draft
\title{
Exact dimer ground state of the two dimensional Heisenberg
spin system SrCu$_2$(BO$_3$)$_2$
}
\author{
Shin Miyahara and Kazuo Ueda
}
\address{Institute for Solid State Physics, University of Tokyo,\\
7-22-1 Roppongi, Minato-ku, Tokyo 106-8666, Japan
}
\date{\today}

\maketitle

\begin{abstract}
The two dimensional Heisenberg model for SrCu$_2$(BO$_3$)$_2$ 
has the exact dimer ground state which was proven by Shastry 
and Sutherland almost twenty years ago. 
The critical value of the   
quantum phase transition from the dimer state to the 
N\'{e}el ordered state is determined.   
Analysis of the experimental data shows 
that SrCu$_2$(BO$_3$)$_2$ has the dimer ground state 
but is close to the transition point, which leads to 
the unusual temperature dependence of the susceptibility. 
Almost localized nature of the triplet excitations explains
the plateaus observed in the magnetization curve.

\end{abstract}



\narrowtext
The pseudo spin-gap behaviors observed in high $T_c$ cuprates have 
stimulated intensive investigations on magnetic systems with spin gaps.
As products 
of this type of activities, several new spin gap systems have been found 
experimentally. Among them, some of the examples which have two 
dimensional character include the coupled spin ladder 
systems, SrCu$_2$O$_3$\cite{azuma}, CaV$_2$O$_5$\cite{iwase} 
and the plaquette RVB system, CaV$_4$O$_9$\cite{sato}. 

Recently Kageyama et al found a new two dimensional spin gap system 
SrCu$_2$(BO$_3$)$_2$\cite{kageyama}.  
The crystal structure of SrCu$_2$(BO$_3$)$_2$
is tetragonal and all Cu$^{2+}$ ions with a localized spin $S=1/2$ are
located at crystallographically equivalent sites. The two dimensional 
layers containing the Cu$^{2+}$ ions are separated by planes of
Sr$^{2+}$ ions.  

The susceptibility drops sharply below the maximum
at around $T=20\ {\rm K}$.    
The peak of the measured susceptibility is much suppressed compared 
with that of the theoretical value expected for a dimer model. 
The spin gap estimated by the nuclear
magnetic relaxation rate is $\Delta=30\ {\rm K}$.  
Another novel feature appears in the magnetization under
high magnetic fields. The authors report that the two 
magnetization plateaus corresponding to 1/4 and 1/8 of the full 
Cu moment are observed for the first time for the two-dimensional 
quantum spin systems. 

\begin{figure}
\psbox[width=7.5cm]{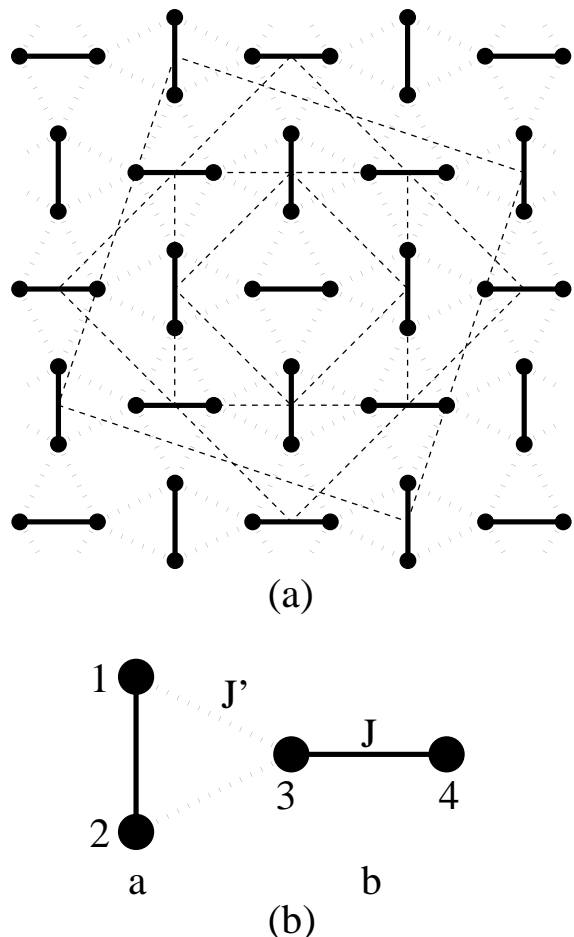}
\caption{
(a) Lattice structure of the Cu$^{2+}$ spins of SrCu$_2$(BO$_3$)$_2$. The
nearest-neighbor bonds are expressed by solid lines and the
next-nearest-neighbor bonds by broken lines. Square unit cells containing
4, 8, 16, 20 spins are shown by dashed lines. (b) Elementary unit
for the interaction between a pair of nearest-neighbor bonds.
}
\label{fig:lattice}
\end{figure}

The magnetic properties of SrCu$_2$(BO$_3$)$_2$ may be
described by the two-dimensional Heisenberg model 
with the nearest-neighbor and next-nearest-neighbor couplings: 
\begin{equation}
  {\cal H} =J\sum_{\rm n.n.} {\bf s}_i \cdot {\bf s}_j
           +J~{'}\sum_{\rm n.n.n.} {\bf s}_i \cdot {\bf s}_j\ .
\label{model}
\end{equation}
The system is illustrated in Fig.\ref{fig:lattice}(a) and 
topologically equivalent to the model considered by Shastry and 
Sutherland \cite{shastry}.
We note that the nearest-neighbor bonds define a unique covering 
of all spins.  On the other hand, the system with only
next-nearest-neighbor couplings is equivalent to
the square lattice Heisenberg model.  
We assume that the both coupling constants are antiferromagnetic,
$J\ {\rm and}\ J^{'} > 0$, and then the present Heisenberg model
is frustrated. 

Shastry and Sutherland have shown that the singlet dimer state 
is an exact eigenstate of the Hamiltonian. 
In this paper we determine the critical value for the 
transition from the gapful dimer ground state 
to the antiferromagnetically
ordered gapless state.  Analysis 
of experimental data shows that SrCu$_2$(BO$_3$)$_2$ is a spin 
gap system with the dimer ground state but is close to this  
quantum transition point.  The unusual magnetic properties of  
SrCu$_2$(BO$_3$)$_2$ are explained consistently from this point 
of view. 

In the following analysis we use the dimer bases defined for 
each nearest-neighbor bond:
\begin{eqnarray}
   | s\rangle & =&\frac{1}{\sqrt{2}}(| \uparrow \downarrow \rangle -
        |  \downarrow \uparrow \rangle ), \\
   | t_1 \rangle & =& | \uparrow \uparrow \rangle, \\
   | t_0 \rangle & =&\frac{1}{\sqrt{2}}(| \uparrow \downarrow \rangle +
        |  \downarrow \uparrow \rangle ), \\
   | t_{-1} \rangle & = &| \downarrow \downarrow \rangle.
\end{eqnarray}
As Shastry and Sutherland\cite{shastry} have shown,
the direct product of the singlet states
\begin{equation}
   |\Psi\rangle = \prod_a | s\rangle_a\ ,
\end{equation}
where $a$ represents the nearest-neighbor bonds,
is always an eigenstate of the Hamiltonian (\ref{model}). 

The proof for the exact eigenstate is simple.  Since the wave
function is an eigenstate, actually the ground state, of the 
first term of the Hamiltonian, let us consider the effect of
the second term.  It is easy to show by elementary 
calculations that for any neighboring pair of the nearest-neighbor 
bonds the matrix element of the second term of Eq.(\ref{model}) vanishes: 
\begin{equation}
  {\cal H^{'}}_{ab}  | s\rangle_a | s\rangle_b = 0.
\end{equation}
To be explicit, ${\cal H^{'}}_{ab}=J^{'}({\bf s}_1\cdot 
{\bf s}_3+{\bf s}_2 \cdot {\bf s}_3)$ where the site indices 
are shown in Fig.\ref{fig:lattice}(b).  Note that the 
vanishing of the matrix element is due to the odd parity of the
singlet with respect to the reflection  which exchanges the 
two spins ${\bf s}_1 \leftrightarrow {\bf s}_2$. 

The energy of the dimer state is given by $E_{\rm dimer} = - (3/8) N J$ 
where $N$ is the total number of spins which is assumed to be even.
It is clear that the dimer state is the ground state for small 
$ J^{'} \ll J$.

Now we consider the other limit, $J^{'} \gg J$.  In this limit the 
model is topologically equivalent to the two dimensional square 
lattice Heisenberg model as mentioned before.  
According to the recent quantum Monte Carlo simulations the 
ground state energy per site is $-0.669 J^{'}$ \cite{sandvik}. The 
first order correction due to the dimer coupling $J$ is obtained from
the spin-spin correlation of the next-nearest-neighbor pair 
of the square lattice model, which is calculated as $0.204$
\cite{troyer}.  By using these results the ground state energy 
of the ordered phase is estimated as  $E_{AF}= -N (0.669 J^{'} 
- 0.102 J$).  The transition point between the two phases may 
be obtained by equating the two energies, which leads to 
$(J^{'}/J)_c = 0.71$. 
Since the ground state energy for the antiferromagnetic phase 
is estimated by a variational calculation,
the transition point thus obtained gives an upper bound.

\begin{figure}
\psbox[width=8.5cm]{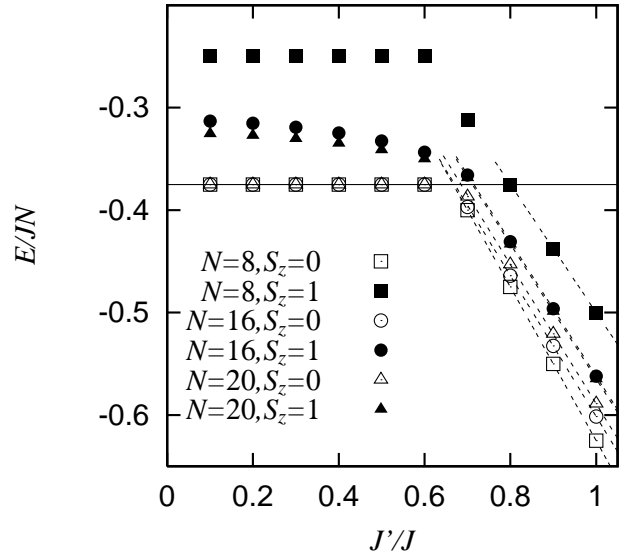}
\caption{
Ground state energy per site for finite lattices: $N=8,\ 16,\ 20$.
The lowest energies of triplet excitations are also shown by filled symbols.
}
\label{fig:energy}
\end{figure}

Figure \ref{fig:energy} shows the energy obtained by the exact 
diagonalization for finite systems, $N=8$, $N=16$ and $N=20$.  
Periodic boundary conditions are used for the calculations.
By considering the finite 
size effects,  we obtain an estimation, $(J^{'}/J)_c > 0.69$.  
Therefore we conclude that the transition point
is $(J^{'}/J)_c = 0.7 \pm 0.01$.

One can estimate the excitation gap in the dimer phase by the
perturbation theory.  After some calculations the spin 
gap up to the fourth order correction is given by
\begin{equation}
  \Delta = J\left( 1 - (\frac{J^{'}}{J})^2 -
  \frac{1}{2}(\frac{J^{'}}{J})^3 - \frac{1}{8}(\frac{J^{'}}{J})^4
  \right).
\label{gap}
\end{equation}
Up to this order the triplet excitations are completely localized.
This unusual behavior is understood by considering the matrix 
elements for the triplet excitations.  To be explicit we start 
from the state with a triplet with the spin quantum number $s_z=1$ 
at the bond $a$ in Fig.\ref{fig:lattice}(b).  
The states connected with this state by the Hamiltonian are 
\begin{equation}
  {\cal H^{'}}_{ab}  | t_1\rangle_a | s\rangle_b = 
  \frac{J^{'}}{2}| t_1\rangle_a | t_0\rangle_b
  - \frac{J^{'}}{2}| t_0\rangle_a | t_1\rangle_b\ .
\label{trail}
\end{equation}   
It is important to note that when a triplet moves to neighboring 
bonds it leaves another triplet behind.  Next crucial observation
is that the following matrix elements vanish also by the symmetry 
reason : parity with respect to the reflection,
\begin{equation}
  {\cal H^{'}}_{ab}  | s\rangle_a | t_m\rangle_b  =0 \ \ \ \ \ \  
 (m=0,\pm1) .
\label{singlet}
\end{equation}
The above two facts, Eq.(\ref{trail}) and (\ref{singlet}), set a 
stringent constraint for motion of a triplet.  Hopping of a triplet 
is possible through a closed path of dimer bonds and thus the hopping
processes start from the sixth order in the perturbation.  On the 
other hand, in a similar system in one dimension, the triplet excitations
are completely localized \cite{gelfand}.  In the present system, the
triplet excitations are nearly localized with extremely small dispersion. 

\begin{figure}
\psbox[width=8.5cm]{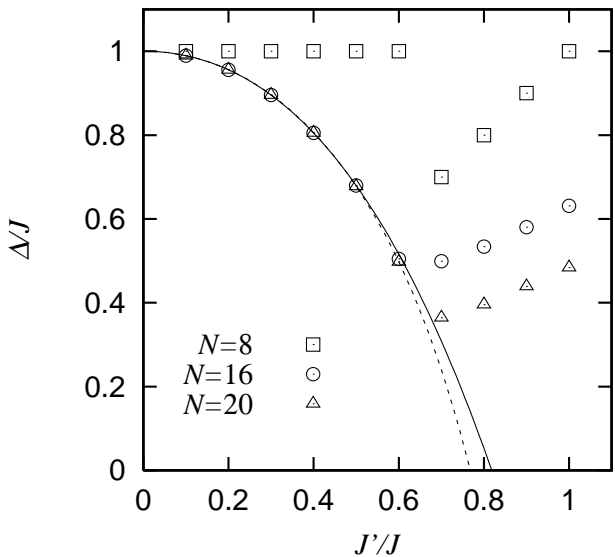}
\caption{
Spin gap for finite lattices: $N=8,\ 16,\ 20$.  The solid line 
is the perturbation result up to the fourth order.  The dashed line 
is a fit obtained by adding fifth and sixth order terms to the 
perturbation result. 
}
\label{fig:gap}
\end{figure}

The spin gap for finite systems is shown in Fig.\ref{fig:gap}.
For the dimer region, $J^{'}/J < 0.7$, the finite size effects are small,
which is a consequence of the almost localized wave functions of the 
triplet excitations. The perturbation result given by Eq.(\ref{gap})
is very accurate for $J^{'}/J \leq 0.5$. 
On the other hand for the antiferromagnetic region
the finite size effects are significant, indicating usual dispersive 
magnon excitations. Since the dimer ground state is always an 
eigenstate it makes a level crossing with the N\'{e}el ordered state
at the transition point. For such a case there are two possibilities
concerning the nature of the transition \cite{affleck}. From the
results shown in Fig. 3 it is difficult to determine uniquely whether
it is weakly first order or continuous.

Susceptibility at high temperatures is obtained by the
high-temperature series expansion as
\begin{equation}
  \chi = \frac{(g\mu_B)^2}{4T}\left( 1 - \frac{J+4J^{'}}{4T}
       + \frac{-J^2+8JJ^{'}+8{J^{'}}^2}{16T^2} \right).
\label{highT}
\end{equation}
From the expansion, the paramagnetic Weiss constant is given by $\theta = 
(J+4J^{'})/4$.

\begin{figure}
\psbox[width=8.5cm]{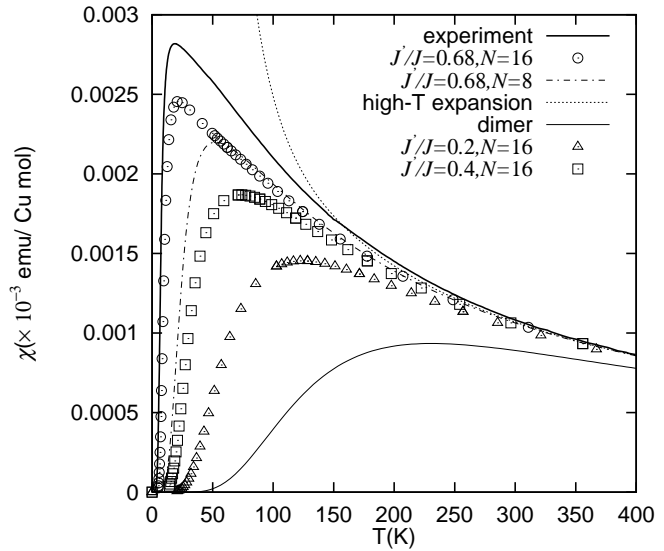}
\caption{
Temperature dependence of susceptibility for finite clusters,
$N=8$ and $16$. $J=100  \ {\rm K}$ and $J^{'}=68 \ {\rm K}$ are 
used.  The experimental data are shown by the thick solid line.
Also shown are the theoretical results
for smaller values of the ratio $J^{'}/J = 0 {\rm (dimer)}, 0.2, 0.4$ with
the Weiss constant fixed $\theta=92.5  \ {\rm K}$.
}
\label{fig:fit}
\end{figure}

Kageyama et al determined the excitation gap $\Delta = 30 \ {\rm K}$
>from the NMR relaxation rate.  It should be noted that from the 
low temperature increase of the susceptibility it is estimated as
$\Delta = 19 \ {\rm K}$.  In the following analysis of 
the experimental data we use the 
former value which is identified as the spin gap by the authors of 
\cite{kageyama}.  The susceptibility data at high 
temperatures are fitted by a Curie-Weiss law with the
Weiss constant $\theta=92.5  \ {\rm K}$ and the effective $g$-factor 
$g=2.14$.  The spin gap and the Weiss constant are sufficient to 
determine the coupling
constants uniquely.  By using the fit up to the sixth order shown in 
Fig.\ref{fig:gap}, $J=100  \ {\rm K}$ and $J^{'}=68 \ {\rm K}$ are 
obtained. Temperature dependence of susceptibility is shown in 
Fig.\ref{fig:fit}.  In the figure comparison is made between the 
theoretical calculations for finite clusters $N=8$ and $N=16$ and 
the experiments.  The theoretical values are calculated by the 
quantum transfer matrix method for the systems with periodic 
boundary conditions \cite{betsuyaku}.
Considering the ambiguity for the estimation of the gap from 
the experiments on one side, and the smallness of clusters used
for theoretical calculations on the other side, the agreement
between the experiment and the theory is satisfactory.
A better fit may be obtained when slightly smaller values are used
for the Weiss constant and the spin gap.

The characteristic feature of the susceptibility of this 
compound, the usual Curie-Weiss type behavior at higher temperatures
above the peak and the steep drop below the peak, is well reproduced. 
The estimated coupling constants $J^{'}/J =0.68$ is close to the critical
value $(J^{'}/J)_c=0.7$.  In Fig.\ref{fig:fit} also shown 
are the theoretical results
for smaller values of the ratio $J^{'}/J = 0, 0.2, 0.4$ with 
the Weiss constant fixed, $\theta=92.5  \ {\rm K}$.  
We can conclude that the closeness to the transition
point is the origin of the unusual temperature dependence 
of the susceptibility. 

\begin{figure}
\psbox[width=8.5cm]{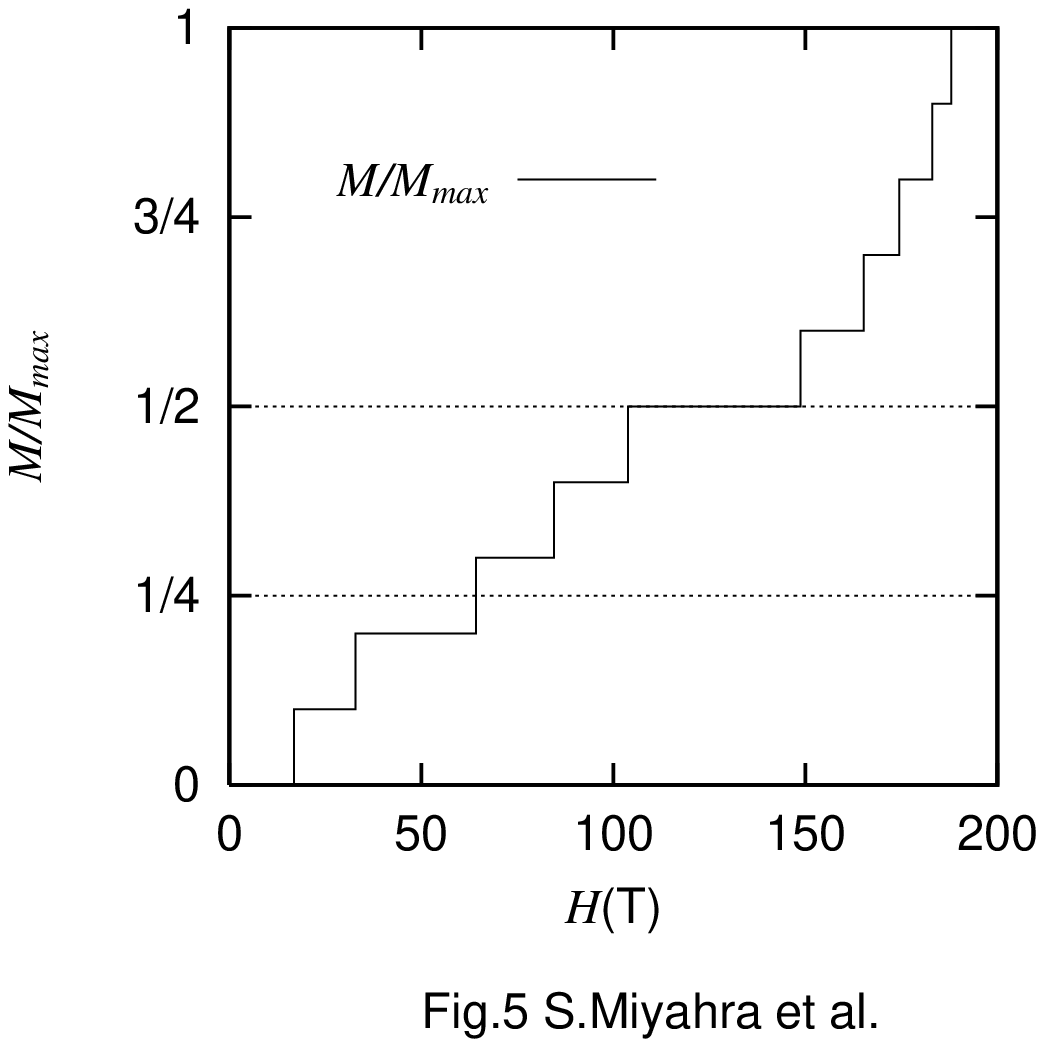}
\caption{
Magnetization process for the finite cluster of $20$ spins. 
$J=100  \ {\rm K}$ and $J^{'}=68 \ {\rm K}$ are 
used.  
}
\label{fig:magnetization}
\end{figure}

Next we discuss the magnetization curve.  Figure \ref{fig:magnetization}
shows magnetization as a function of applied magnetic field. The
results are obtained by the numerical exact diagonalization for
the system with  $N=20$ spins. 
One can clearly identify plateaus corresponding to 1/4 and 1/2 of the 
full moment.  Experimentally, plateaus are observed for 1/4 and 
1/8 of the full moment. Since the calculations were done for 
a small cluster, it is not possible to find 
a plateau at 1/8.  The critical value for the plateau at 1/4
estimated for the present set of parameters $J=100  \ {\rm K}$, $J^{'}=68 
\ {\rm K}$ is  about 40 T which is comparable with the experimentally
observed critical field $H_{c4}= 37 {\rm T}$\cite{kageyama}.  

Formation of the plateaus are related with the almost localized wave 
functions of the triplet excitations.  At special values of magnetization 
where the triplet excitations can take a regular lattice structure, 
energy cost would be locally minimum. 
It is natural to assume that the commensurability energy is
more favorable when a unit cell is a simple square
since the original structure has the tetragonal symmetry.  
The square unit cells are possible
for $N=4,\ 8,\ 16,\ 20,\ 32, \cdots$  
which corresponds to plateaus at $1/2,\ 1/4,\ 1/8,\ 1/10,\
1/16, \cdots$, see Fig.\ref{fig:lattice}(a). 
The plateaus observed experimentally, 1/4 and 1/8, are 
just two of them.  Our theory predicts that there will be a clear plateau
for 1/2 of the full moment at around 100 T and 
also for 1/10 at a smaller magnetic field than 
$H_{c2}=28 {\rm T}$.  Since the plateau of the latter is expected 
to be small, a lower temperature and a clean sample will be required.
The experiments on these plateaus are a crucial test for the present
theory. Another prediction of the theory is the superstructure
at each plateau which should be observable, for example by NMR.
  
In conclusion we have identified that SrCu$_2$(BO$_3$)$_2$ has 
the exact dimer ground state found by Shastry and Sutherland
\cite{shastry}.
In the model the quantum 
phase transition occurs at $J^{'}/J\sim 0.7$ from the dimer
state to the antiferromagnetically ordered state.
This transition is either weakly first order or continuous
and SrCu$_2$(BO$_3$)$_2$ is a 
unique spin gap system which is close to the quantum transition point.
Unusual temperature dependence of the susceptibility is a 
consequence of the closeness to the transition point.
Another interesting feature appears in the excitations. 
The low lying triplet excitations are almost localized and 
easy to form regular lattices under certain magnetic fields.
The commensurability energy associated with the superstructures
leads to the plateaus 
in magnetization curve at $1/2,\ 1/4,\ 1/8,\ 1/10,\ 1/16, 
\cdots$ of the full moment. 

We would like to express our sincere thanks to Dr. H. Kageyama 
for showing us his experimental results. 
We are also grateful to Matthias Troyer
for providing the result of Quantum Monte Carlo simulations for the square 
lattice Heisenberg model.
After the submission of the original version of the present paper,
we were informed of the work by Shastry and Sutherland from Dr.
L.O. Manuel to whom we are thankful.  
The numerical exact diagonalization calculations were done by using
TITPACK ver.2 developed by H. Nishimori. A part of the numerical
calculations was performed on the HITACHI SR2201 massively parallel
computer of the University of Tokyo.


\begin{thebibliography}{99}

\bibitem{azuma} M. Azuma, Z. Hiroi, M. Takano, K. Ishida, and Y.
Kitaoka, Phys. Rev. Lett. {\bf 73}, 3463 (1994).
\bibitem{iwase} H. Iwase, M. Isobe, Y. Ueda, and H. Yasuoka,
J. Phys. Soc. Jpn {\bf 65}, 2397 (1996).
\bibitem{sato} S. Taniguchi, T. Nishikawa, Y. Yasui, Y. Kobayashi,
 M. Sato, T. Nishioka, M. Kontani, and K. Sano,
 J. Phys. Soc. Jpn {\bf 64}, 2758 (1995).
\bibitem{kageyama} H. Kageyama, K. Yoshimura, R. Stern,
 N.V. Mushnikov, K. Onizuka, M.Kato, K. Kosuge, C.P. Slichter, T. Goto, and
 Y. Ueda, Phys. Rev. Lett.(1999) in press.
\bibitem{shastry} B.S. Shastry and B. Sutherland,
Physica 108{\bf B}, 1069 (1981).
\bibitem{sandvik} A.W. Sandvik, Phys. Rev. B{\bf 56}, 11678 (1997).
\bibitem{troyer} M. Troyer, private communication.
\bibitem{gelfand} M.P. Gelfand, Phys. Rev. B{\bf 43}, 8644 (1991).
\bibitem{affleck} I. Affleck, Phys. Rev. B{\bf 43}, 3215 (1991).
\bibitem{betsuyaku} H. Betsuyaku, Prog. Theor. Phys. {\bf 75},
774 (1986); I. Morgenstern and D. W\"{u}rtz, Phys. Rev. B {\bf 32},
532 (1985).

\end{thebibliography}
\end{document}